\documentclass[
journal=jctcce, % for undefined journal
manuscript=article]{achemso}

\usepackage[version=3]{mhchem} % Formula subscripts using \ce{}
\usepackage{natbib}
\usepackage{graphicx}
\usepackage{subcaption}
\usepackage{longtable} 
\usepackage{notes2bib}
\usepackage{url}
\usepackage{array}
\usepackage{hyperref}
\usepackage{multirow}
\usepackage{marginnote}
\usepackage{bm}
\usepackage{physics}
\usepackage{braket}

\title{A Finite-field Approach for $GW$ Calculations Beyond the Random Phase Approximation}

\author{He Ma}
\affiliation
{Institute for Molecular Engineering, University of Chicago, Chicago, Illinois 60637, United States}
\alsoaffiliation
{Department of Chemistry, University of Chicago, Chicago, Illinois 60637, United States}

\author{Marco Govoni}
\affiliation
{Institute for Molecular Engineering, University of Chicago, Chicago, Illinois 60637, United States}
\alsoaffiliation
{Materials Science Division, Argonne National Laboratory, Lemont, Illinois 60439, United States.}

\author{Francois Gygi}
\affiliation
{Department of Computer Science, University of California Davis, Davis, California 95616, United States.}

\author{Giulia Galli}
\affiliation
{Institute for Molecular Engineering, University of Chicago, Chicago, Illinois 60637, United States}
\alsoaffiliation
{Department of Chemistry, University of Chicago, Chicago, Illinois 60637, United States}
\alsoaffiliation[MSD]
{Materials Science Division, Argonne National Laboratory, Lemont, Illinois 60439, United States.}
\email{gagalli@uchicago.edu}

\begin{document}

\begin{abstract}
We describe a finite-field approach to compute density response functions, which allows for efficient $G_0W_0$ and $G_0W_0\Gamma_0$ calculations beyond the random phase approximation. The method is easily applicable to density functional calculations performed with hybrid functionals. We present results for the electronic properties of molecules and solids and we discuss a general scheme to overcome slow convergence of quasiparticle energies obtained from $G_0W_0\Gamma_0$ calculations, as a function of the basis set used to represent the dielectric matrix.
\end{abstract}

\section{Introduction}

Accurate, first principles predictions of the electronic structure of molecules and materials are important goals in chemistry, condensed matter physics and materials science \cite{Onida2002}. In the past three decades, density functional theory (DFT) \cite{Hohenberg1964, Kohn1965} has been successfully adopted to predict numerous properties of molecules and materials \cite{Becke2014}. In principle, any ground or excited state properties can be formulated as functionals of the ground state charge density. In practical calculations, the ground state charge density is determined by solving the Kohn-Sham (KS) equations with approximate exchange-correlation functionals, and many important excited state properties are not directly accessible from the solution of the KS equations. The time-dependent formulation of DFT (TDDFT) \cite{Runge1984} in the frequency domain \cite{Casida1995} provides a computationally tractable method to compute excitation energies and absorption spectra. However, using the common adiabatic approximation to the exchange-correlation functional, TDDFT is often not sufficiently accurate to describe certain types of excited states such as Rydberg and charge transfer states \cite{Casida2012}, especially when semilocal functionals are used.

A promising approach to predict excited state properties of molecules and materials is many-body perturbation theory (MBPT) \cite{Hedin1965, Hybertson1986, Martin2016}. Within MBPT, the $GW$ approximation can be used to compute quasiparticle energies that correspond to photoemission and inverse photoemission measurements; furthermore, by solving the Bethe-Salpeter equation (BSE), one can obtain neutral excitation energies corresponding to optical spectra. For many years since the first applications of MBPT \cite{Hybertson1986}, its use has been hindered by its high computational cost. In the last decade, several advances have been proposed to improve the efficiency of MBPT calculations \cite{Umari2009, Neuhauser2014, Liu2016}, which are now applicable to simulations of relatively large and complex systems, including nanostructures and heterogeneous interfaces \cite{Ping2013, Pham2017, Leng2016}. In particular, $GW$ and BSE calculations can be performed using a low rank representation of density response functions \cite{Nguyen2012, Pham2013, Govoni2015, Govoni2018}, whose spectral decomposition is obtained through iterative diagonalization using density functional perturbation theory (DFPT) \cite{Baroni1987, Baroni2001}. This method does not require the explicit calculation of empty electronic states and avoids the inversion or storage of large dielectric matrices. The resulting implementation in the WEST code \bibnote{WEST. http://www.west-code.org/ (accessed  Aug. 1, 2018).} has been successfully applied to investigate numerous systems including defects in semiconductors \cite{Seo2016, Seo2017}, nanoparticles\cite{Scherpelz2016}, aqueous solutions\cite{Gaiduk2016,Pham2017,Gaiduk2018}, and solid/liquid interfaces\cite{Govoni2015,Gerosa2018} .

In this work, we developed a finite-field (FF) approach to evaluate density response functions entering the definition of the screened Coulomb interaction $W$. The FF approach can be used as an alternative to DFPT, and presents the additional advantage of being applicable, in a straightforward manner, to both semilocal and hybrid functionals. In addition, FF calculations allow for the direct evaluation of density response functions beyond the random phase approximation (RPA).

Here we first benchmark the accuracy of the FF approach for the calculation of several density response functions, from which one can obtain the exchange correlation kernel ($f_{\text{xc}}$), defined as the functional derivative of the exchange-correlation potential with respect to the charge density. Then we discuss $G_0W_0$ calculations for various molecules and solids, carried out with either semilocal or hybrid functionals, and by adopting different approximations to include vertex corrections in the self-energy. In the last two decades a variety of methods \cite{DelSole1994, Fleszar1997, Schindlmayr1998, Marini2004, Bruneval2005, Tiago2006, Morris2007, Shishkin2007, Shaltaf2008, Romaniello2009, Gruneis2014, Chen2015, Kutepov2016, Kutepov2017, Maggio2017} \bibnote{Lewis, A. M.; Berkelbach, T. C. Vertex corrections to the polarizability do not improve the GW approximation for molecules. 2004, arXiv:1810.00456. arXiv.org ePrint archive. http://arxiv.org/abs/1810.00456 (accessed Oct 1, 2018).} has been proposed to carry out vertex-corrected $GW$ calculations, with different approximations to the vertex function $\Gamma$ and including various levels of self-consistency between $G$, $W$ and $\Gamma$. Here we focus on two formulations that are computationally tractable also for relatively large systems, denoted as $G_0W_0^{f_\text{xc}}$ and $G_0W_0\Gamma_0$. In $G_0W_0^{f_\text{xc}}$, $f_{\text{xc}}$ is included in the evaluation of the screened Coulomb interaction $W$; in $G_0W_0\Gamma_0$, $f_{\text{xc}}$ is included in the calculation of both $W$ and the self-energy $\Sigma$ through the definition of a local vertex function. Most previous $G_0W_0^{f_\text{xc}}$ and $G_0W_0\Gamma_0$ calculations were restricted to the use of the LDA functional \cite{DelSole1994, Fleszar1997, Tiago2006, Morris2007}, for which an analytical expression of $f_{\text{xc}}$ is available. Paier \textit{et al.} \cite{Paier2008} reported $GW_0^{f_\text{xc}}$ results for solids obtained with the HSE03 range-separated hybrid functional \cite{Heyd2003}, and the exact exchange part of $f_{\text{xc}}$ is defined using the nanoquanta kernel \cite{Reining2002, Marini2003, Sottile2003, Bruneval2005}. In this work semilocal and hybrid functionals are treated on equal footing, and we present calculations using LDA \cite{Perdew1981}, PBE \cite{Perdew1997} and PBE0 \cite{Perdew1996} functionals, as well as a dielectric-dependent hybrid (DDH) functional for solids \cite{Skone2014}. 

A recent study of Thygesen and co-workers \cite{Schmidt2017} reported basis set convergence issues when performing $G_0W_0\Gamma_0@\text{LDA}$ calculations, which could be overcome by applying a proper renormalization to the short-range component of $f_{\text{xc}}$ \cite{Olsen2012, Olsen2013, Patrick2015}. In our work we generalized the renormalization scheme of Thygesen \textit{et al.} to functionals other than LDA, and we show that the convergence of $G_0W_0\Gamma_0$ quasiparticle energies is significantly improved using the renormalized $f_\text{xc}$.

The rest of the paper is organized as follows. In Section 2 we describe the finite-field approach and benchmark its accuracy. In Section 3 we describe the formalism used to perform $GW$ calculations beyond the RPA, including a renormalization scheme for $f_{\text{xc}}$, and we compare the quasiparticle energies obtained from different $GW$ approximations (RPA or vertex-corrected) for molecules in the GW100 test set \cite{vanSetten2015} and for several solids. Finally, we summarize our results in Section 4.

\section{The finite-field approach}

We first describe the FF approach for iterative diagonalization of density response functions and we then discuss its robustness and accuracy.

\subsection{Formalism}

Our $G_0W_0$ calculations are based on DFT single-particle energies and wavefunctions, obtained by solving the Kohn-Sham (KS) equations: 
\begin{equation} \label{KS}
    H_{\text{KS}} \psi_{m}(\bm{r}) = \varepsilon_{m} \psi_{m}(\bm{r}),
\end{equation}
where the KS Hamiltonian $H_{\text{KS}} = T + V_{\text{SCF}} = T + V_{\text{ion}} + V_{\text{H}} + V_{\text{xc}}$. $T$ is the kinetic energy operator; $V_{\text{SCF}}$ is the KS potential that includes the ionic $V_{\text{ion}}$, the Hartree $V_{\text{H}}$ and the exchange-correlation potential $V_{\text{xc}}$. The charge density is given by $n(\bm{r}) = \sum_m^{\text{occ.}} \left| \psi_{m}(\bm{r}) \right|^2$. For simplicity we omitted the spin index.

We consider the density response function (polarizability) of the KS system $\chi_0(\bm{r}, \bm{r}')$ and that of the physical system $\chi(\bm{r}, \bm{r}')$; the latter is denoted as $\chi_{\text{RPA}}(\bm{r}, \bm{r}')$ when the random phase approximation (RPA) is used. The variation of the charge density due to either a variation of the KS potential $\delta V_{\text{SCF}}$ or the external potential $\delta V_{\text{ext}}$ is given by:
\begin{equation} \label{kernel}
    \delta n(\bm{r}) = \int K(\bm{r}, \bm{r}') \delta V(\bm{r}') d\bm{r}',
\end{equation}
where $K = \chi_0(\bm{r}, \bm{r}')$ if $\delta V(\bm{r}') = \delta V_{\text{SCF}}(\bm{r}')$ and $K = \chi(\bm{r}, \bm {r'})$ if $\delta V(\bm{r}') = \delta V_{\text{ext}}(\bm{r}')$. The density response functions of the KS and physical system are related by a Dyson-like equation:  
\begin{equation} \label{Dyson}
    \chi(\bm{r}, \bm{r}') = \chi_{0}(\bm{r}, \bm{r}') + \int d\bm{r}'' \int d\bm{r}''' \chi_{0}(\bm{r}, \bm{r}'') \left[v_{\text{c}}(\bm{r}'', \bm{r}''') + f_{\text{xc}}(\bm{r}'', \bm{r}''')\right] \chi(\bm{r}''', \bm{r}')
\end{equation}
where $v_{\text{c}}(\bm{r}, \bm{r}') = \frac{1}{|\bm{r} - \bm{r}'|}$ is the Coulomb kernel and $f_{\text{xc}}(\bm{r}, \bm{r}') = \frac{\delta V_{\text{xc}}(\bm{r})}{\delta n(\bm{r}')}$ is the exchange-correlation kernel.

Within the RPA, $f_{\text{xc}}$ is neglected and $\chi(\bm{r}, \bm{r}')$ is approximated by:
\begin{equation} \label{DysonRPA}
    \chi_{\text{RPA}}(\bm{r}, \bm{r}') = \chi_{0}(\bm{r}, \bm{r}') + \int d\bm{r}'' \int d\bm{r}''' \chi_{0}(\bm{r}, \bm{r}'') v_{\text{c}}(\bm{r}'', \bm{r}''') \chi(\bm{r}''', \bm{r}').
\end{equation}

In the plane-wave representation (for simplicity we only focus on the $\Gamma$ point of the Brillouin zone), $v_{\text{c}}(\bm{G}, \bm{G}') = \frac{4\pi \delta(\bm{G}, \bm{G}')}{|\bm{G}|^2}$ (abbreviated as $v_{\text{c}}(\bm{G}) = \frac{4\pi}{|\bm{G}|^2}$). We use $K(\bm{G}, \bm{G}')$ to denote a general response function ($K \in \{ \chi_0, \chi_\text{RPA}, \chi \}$), and define the dimensionless response function $\tilde{K}(\bm{G}, \bm{G}')$ ($\tilde{K} \in \{ \tilde{\chi}_0, \tilde{\chi}_\text{RPA}, \tilde{\chi} \}$) by symmetrizing $K(\bm{G}, \bm{G}')$ with respect to $v_{\text{c}}$:
\begin{equation}
    \tilde{K}(\bm{G}, \bm{G}') = v_{\text{c}}^{\frac{1}{2}}(\bm{G}) K(\bm{G}, \bm{G}') v_{\text{c}}^{\frac{1}{2}}(\bm{G'}).
\end{equation}

The dimensionless response functions $\tilde{\chi}_{\text{RPA}}$ and $\tilde{\chi}_0$ (see eq \ref{DysonRPA}) have the same eigenvectors, and their eigenvalues are related by:
\begin{equation} \label{lambdarpa}
    \lambda_i^{\text{RPA}} = \frac{\lambda_i^0}{1-\lambda_i^0}
\end{equation}
where $\lambda_i^{\text{RPA}}$ and $\lambda_i^0$ are eigenvalues of $\tilde{\chi}_{\text{RPA}}$ and $\tilde{\chi}_0$, respectively. In general the eiegenvalues and eigenvectors of $\tilde{\chi}_{\text{RPA}}$ are different from those of $\tilde{\chi}$ due to the presence of $f_{\text{xc}}$ in eq \ref{Dyson}.

In our \textit{GW} calculations we use a low rank decomposition of $\tilde{K}$:
\begin{equation} \label{pdep}
    \tilde{K} = \sum_i^{N_{\text{PDEP}}} \lambda_i \ket{\xi_i} \bra{\xi_i}
\end{equation}
where $\lambda$ and $\ket{\xi}$ denote eigenvalue and eigenvectors of $\tilde{K}$, respectively. The set of $\xi$ constitute a projective dielectric eigenpotential (PDEP) basis \cite{Nguyen2012, Pham2013, Govoni2015}, and the accuracy of the low rank decomposition is controlled by $N_{\text{PDEP}}$, the size of the basis. In the limit of $N_{\text{PDEP}} = N_{\text{PW}}$ (the number of plane waves), the PDEP basis and the plane wave basis are related by a unitary transformation. In practical calculations it was shown that \cite{Nguyen2012, Pham2013} one only need $N_{\text{PDEP}} \ll N_{\text{PW}}$ to converge the computed quasiparticle energies. To obtain the PDEP basis, an iterative diagonalization is performed for $\tilde{K}$, e.g. with the Davidson algorithm \cite{Davidson1975}. The iterative diagonalization requires evaluating the action of $\tilde{K}$ on an arbitrary trial function $\xi$:
\begin{equation} \label{operation}
\begin{split}
    (\tilde{K}\xi)(\bm{G})
    &= \sum_{\bm{G}'} v_{\text{c}}^{\frac{1}{2}}(\bm{G}) K(\bm{G}, \bm{G}') v_{\text{c}}^{\frac{1}{2}}(\bm{G}') \xi(\bm{G}') \\
    &= v_{\text{c}}^{\frac{1}{2}}(\bm{G}) \mathcal{FT} \left\{ \int K(\bm{r}, \bm{r'}) \left( \mathcal{FT}^{-1} \left[ v_{\text{c}}^{\frac{1}{2}}(\bm{G}') \xi(\bm{G}') \right] \right) (\bm{r}') d\bm{r}' \right\}(\bm{G})
\end{split}
\end{equation}
where $\mathcal{FT}$ and $\mathcal{FT}^{-1}$ denote forward and inverse Fourier transforms respectively. By using eq \ref{operation} we cast the evaluation of $\tilde{K}\xi$ to an integral in real space.

Defining a perturbation $\delta V(\bm{G}')$ = $v_{\text{c}}^{\frac{1}{2}}(\bm{G'}) \xi(\bm{G}')$, the calculation of the real space integral in eq \ref{operation} is equivalent to solving for the variation of the charge density $\delta n$ due to $\delta V$:
\begin{equation} \label{koperation}
    \int K(\bm{r}, \bm{r'}) \left( \mathcal{FT}^{-1} \left[ v_{\text{c}}^{\frac{1}{2}}(\bm{G}') \xi(\bm{G}') \right] \right) (\bm{r}') d\bm{r}' = \int K(\bm{r}, \bm{r'}) \delta V(\bm{r}') d\bm{r}' \equiv \delta n(\bm{r}).
\end{equation}

In previous works $\delta n(\bm{r})$ was obtained using DFPT for the case of $K = \chi_0$ \cite{Govoni2015}. In this work we solve eq \ref{koperation} by a finite-field approach. In particular, we perform two SCF calculations under the action of the potentials $\pm \delta V$:
\begin{equation} \label{perturbedKS}
    (H_{\text{KS}} \pm \delta V) \psi_{m}^{\pm}(\bm{r})  = \varepsilon_{m}^{\pm} \psi_{m}^{\pm}(\bm{r}),
\end{equation}
and $\delta n(\bm{r})$ is computed through a finite difference:
\begin{equation} \label{finitediff}
    \delta{n(\bm{r})} = \frac{1}{2}  \left[ \sum_m^{\text{occ.}} \left| \psi_{m}^{+}(\bm{r}) \right|^2 - \sum_m^{\text{occ.}} \left| \psi_{m}^{-}(\bm{r}) \right|^2\right]
\end{equation}

In eq \ref{finitediff} we use a central difference instead of forward/backward difference to increase the numerical accuracy of the computed $\delta{n(\bm{r})}$.

If in the SCF procedure adopted in eq \ref{perturbedKS} all potential terms in the KS Hamiltonian are computed self-consistently, then the solution of eq \ref{finitediff} yields $K = \chi$ (see eq \ref{koperation}). If  $V_{\text{xc}}$ is evaluated for the initial charge density (i.e. $V_{\text{xc}} = V_{\text{xc}}[n_0]$) and kept fixed during the SCF iterations, then the solution of eq \ref{finitediff} yields $K = \chi_{\text{RPA}}$. If both $V_{\text{xc}}$ and $V_{\text{H}}$ are kept fixed, the solution of eq \ref{finitediff} yields $K = \chi_0$.

Unlike DFPT, the finite-field approach adopted here allows for the straightforward calculation of response functions beyond the RPA (i.e. for the calculation of $\chi$ instead of $\chi_0$ or $\chi_{\text{RPA}}$), and it can be readily applied to hybrid functionals for which analytical expressions of $f_{\text{xc}}$ are not available. We note that finite-field calculations with hybrid functionals can easily benefit from any methodological development that reduces the computational complexity of evaluating exact exchange potentials \cite{Gygi2009, Gygi2013, Dawson2015}.

Once the PDEP basis is obtained by iterative diagonalization of $\tilde{\chi}_{0}$ \bibnote{Here, we defined the PDEP basis to be the eigenvectors of $\tilde{\chi}_{0}$. Alternatively, one may first iteratively diagonalize $\tilde{\chi}$ and define its eigenvectors as the PDEP basis. Then $\tilde{\chi}_0$ and $\tilde{f}_{\text{xc}}$ can be evaluated in the space of the eigenvectors of $\tilde{\chi}$. This choice is not further discussed in the paper; we only mention that some comparisons for the quasiparticle energies (at the $G_0W_0^{f_\text{xc}}$ level, see Section 3) of selected molecules obtained using either $\tilde{\chi}_0$ or $\tilde{\chi}$ eigenvectors as the PDEP basis are identical within 0.01 (0.005) eV for the HOMO (LUMO) state.}, the projection of $\tilde{\chi}$ on the PDEP basis can also be performed using the finite-field approach. Then the symmetrized exchange-correlation kernel $\tilde{f}_{\text{xc}} = v_{\text{c}}^{-\frac{1}{2}} f_{\text{xc}} v_{\text{c}}^{-\frac{1}{2}}$ can be computed by inverting the Dyson-like equation (eq \ref{Dyson}):
\begin{equation} \label{fxcmatrix}
    \tilde{f}_{\text{xc}} = {\tilde{\chi}_{0}}^{-1} - \tilde{\chi}^{-1} - 1.
\end{equation}

On the right hand side of eq \ref{fxcmatrix} all matrices are $N_{\text{PDEP}} \times N_{\text{PDEP}}$ and therefore the resulting $\tilde{f}_{\text{xc}}$ is also defined on the PDEP basis. 

When using orbital-dependent functionals such as meta-GGA and hybrid functionals, the $\tilde{f}_{\text{xc}}$ computed from eq \ref{fxcmatrix} should be interpreted with caution. In this case, DFT calculations for $H_{\text{KS}} \pm \delta V$ can be performed using either the optimized effective potential (OEP) or the generalized Kohn-Sham (GKS) scheme. In the OEP scheme, $v_{\text{xc}}$ is local in space and $f_{\text{xc}}(\bm{r}, \bm{r}') = \frac{\delta V_{\text{xc}}(\bm{r})}{\delta n(\bm{r}')}$ depends on $\bm{r}$ and $\bm{r'}$, as in the case of semilocal functionals. In the GKS scheme, $V_{\text{xc}}$ is non-local and $f_{\text{xc}}(\bm{r}, \bm{r}'; \bm{r}'') = \frac{\delta V_{\text{xc}}(\bm{r}, \bm{r}')}{\delta n(\bm{r}'')}$ depends on three position vectors. We expect $\delta n$ to be almost independent of the chosen scheme, whether GKS or OEP, since both methods yield the same result within first order in the charge density \cite{Kummel2008}. We conducted hybrid functional calculations within the GKS scheme, assuming that for every GKS calculation an OEP can be defined yielding the same charge density; with this assumption the $f_{\text{xc}}$ from eq \ref{fxcmatrix} is well defined within the OEP formalism.

\subsection{Implementation and Verification}

\begin{figure}[H]
  \includegraphics[width=7in]{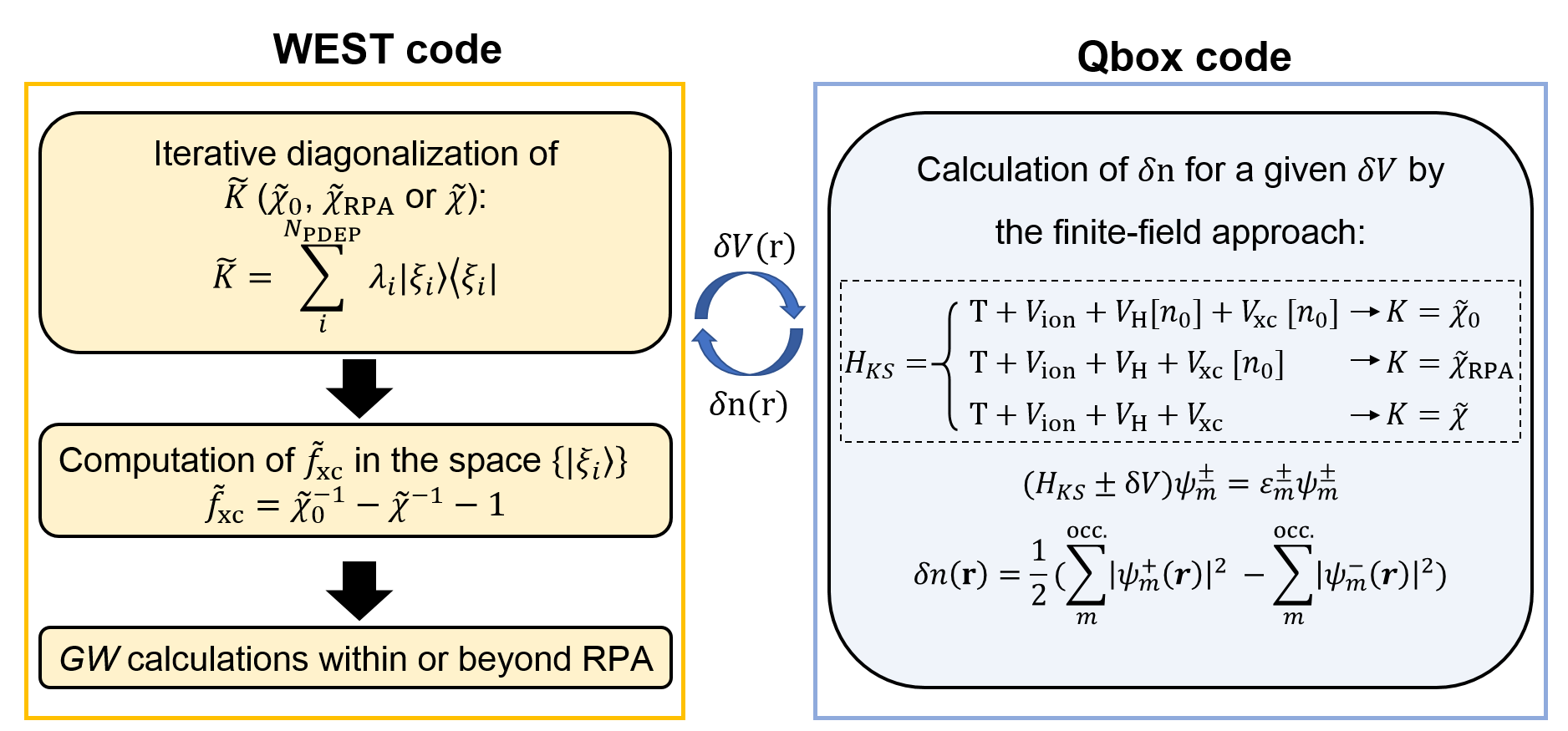}
  \caption{Workflow of finite-field calculations. The WEST code performs an iterative diagonalization of $\tilde{K}$ ($\tilde{\chi}_0$, $\tilde{\chi}_{\text{RPA}}$, $\tilde{\chi}$). In $GW$ calculations beyond the RPA, $\tilde{f}_{\text{xc}}$ is computed from eq \ref{fxcmatrix}, which requires computing the spectral decomposition of $\tilde{\chi}_0$ and evaluating $\tilde{\chi}$ in the space of $\tilde{\chi}_0$ eigenvectors. Finite-field calculations are carried out by the Qbox code. If the Hartree ($V_{\text{H}}$) and exchange correlation potential ($V_{\text{xc}}$) are updated self-consistently when solving eq \ref{perturbedKS}, one obtains $K = \chi$; if  $V_{\text{xc}}$ is evaluated at the initial charge density $n_0$ and kept fixed during the SCF procedure, one obtains $K = \chi_{\text{RPA}}$; if both $V_{\text{xc}}$ and $V_{\text{H}}$ are evaluated for $n_0$ and kept fixed, one obtains $K = \chi_0$. The communications of $\delta n$ and $\delta V$ between WEST and Qbox is carried through the filesystem.}
  \label{qbox_west_coupling}
\end{figure}

We implemented the finite-field algorithm described above by coupling the WEST \cite{Govoni2015} and Qbox \cite{Gygi2008} \bibnote{Qbox. http://www.qboxcode.org (accessed Aug. 1, 2018).} codes in client-server mode, using the workflow summarized in Figure \ref{qbox_west_coupling}. In particular, in our implementation the WEST code performs an iterative diagonalization of $\tilde{K}$ by outsourcing the evaluation of the action of $\tilde{K}$ on an arbitrary function to Qbox, which performs DFT calculations in finite field. The two codes communicate through the filesystem.

To verify the correctness of our implementation, we computed $\tilde{\chi}_0$, $\tilde{\chi}^{\text{RPA}}$, $\tilde{\chi}$ for selected molecules in the GW100 set and we compared the results to those obtained with DFPT. Section 1 of the SI summarizes the parameters used including plane wave cutoff $E_\text{cut}$, $N_{\text{PDEP}}$ and size of the simulation cell. In finite-field calculations we optimized the ground state wavefunction using a preconditioned steepest descent algorithm with Anderson acceleration\cite{Anderson1965}. The magnitude of $\delta V$ was chosen to insure that calculations were performed within the linear response regime (see Section 2 of the SI). All calculations presented in this section were performed with the PBE functional unless otherwise specified.

\begin{figure}[H]
  \includegraphics[width=7in]{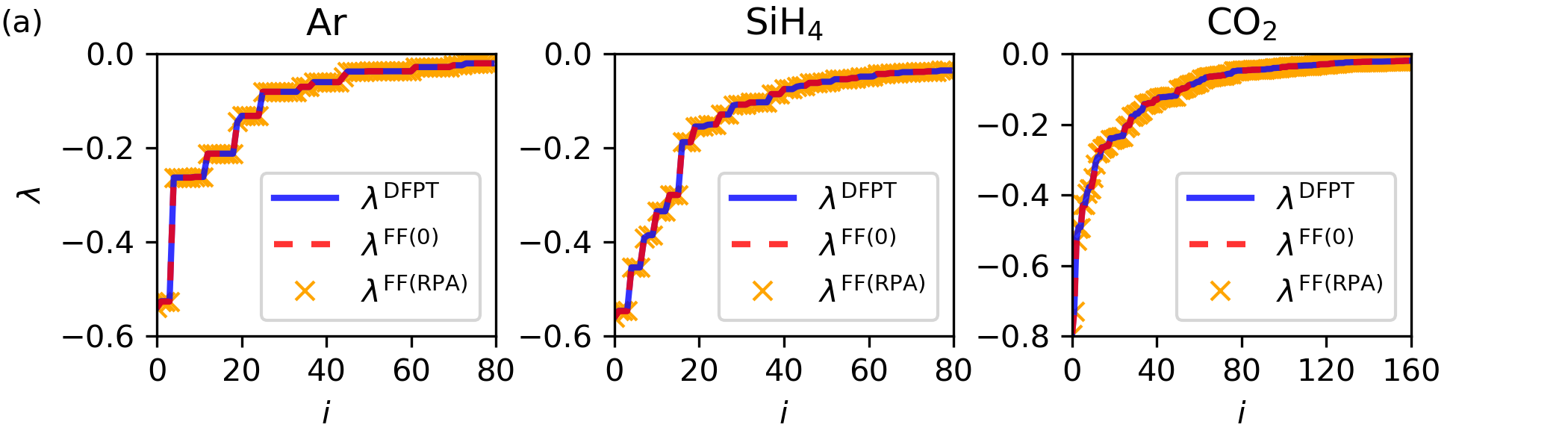}
  \includegraphics[width=7in]{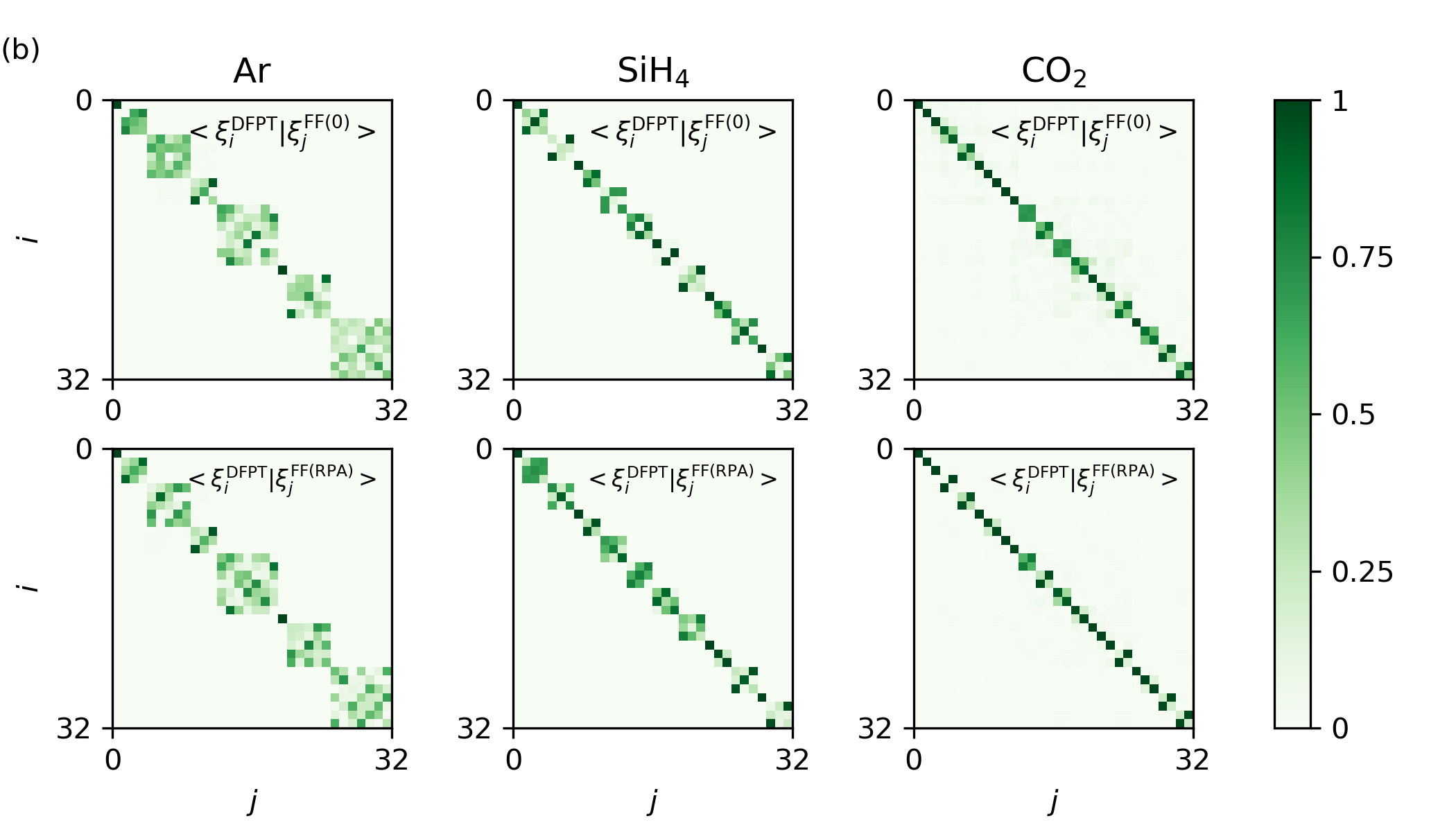}
  \caption{Comparison of the eigenvalues(a) and eigenfunctions(b) of $\tilde{\chi}_{\text{RPA}}$ obtained from density functional perturbation theory (DFPT) and finite-field (FF) calculations. Three approaches are used: diagonalization of $\tilde{\chi}_0$ by DFPT, diagonalization of $\tilde{\chi}_0$ by FF (denoted by FF(0)) and diagonalization of $\tilde{\chi}_{\text{RPA}}$ by FF (denoted by FF(RPA)). In the case of DFPT and FF(0), eq \ref{lambdarpa} was used to obtain the eigenvalues of $\tilde{\chi}_{\text{RPA}}$ from those of $\tilde{\chi}_0$. In (b) we show the first $32 \times 32$ elements of the $\braket{\xi^{\text{DFPT}} | \xi^{\text{FF(0)}}}$ and $\braket{\xi^{\text{DFPT}} | \xi^{\text{FF(RPA)}}}$ matrices (see eq \ref{pdep}). }
  \label{ff_vs_dfpt}
\end{figure}

Figure \ref{ff_vs_dfpt}a shows the eigenvalues of $\tilde{\chi}_{\text{RPA}}$ for a few molecules obtained with three approaches: iterative diagonalization of $\tilde{\chi}_{\text{RPA}}$ with the finite-field approach; iterative diagonalization of $\tilde{\chi}_0$ with either the finite-field approach or with DFPT, followed by a transformation of eigenvalues as in eq \ref{lambdarpa}. The three approaches yield almost identical eigenvalues.

The eigenvectors of the response functions are shown in Figure \ref{ff_vs_dfpt}b, where we report elements of the matrices defined by the overlap between finite-field and DFPT eigenvectors. The inner product matrices are block-diagonal, with blocks corresponding to the presence of degenerate eigenvalues. The agreement between eigenvalues and eigenvectors shown in Figure \ref{ff_vs_dfpt} verifies the accuracy and robustness of finite-field calculations.

\begin{figure}[H]
  \includegraphics[width=7in]{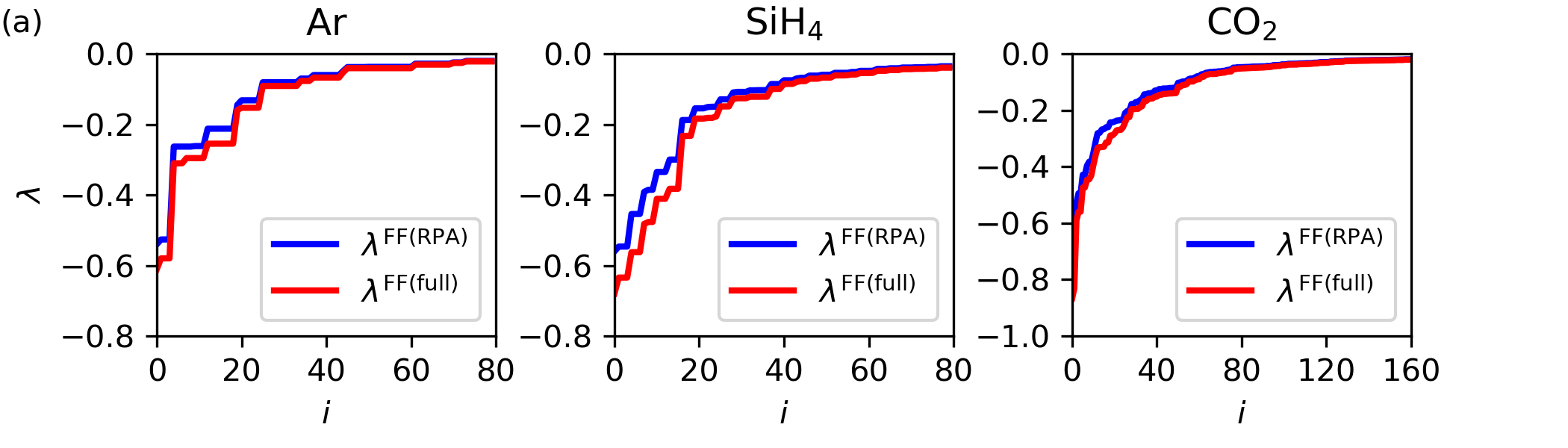}
  \includegraphics[width=7in]{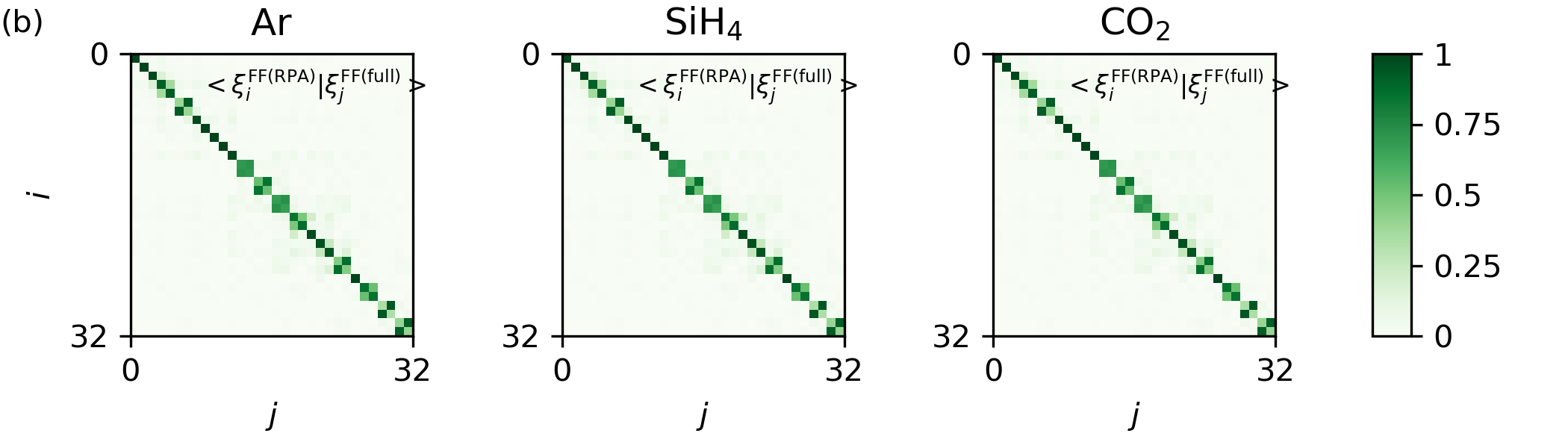}
  \caption{Comparison of eigenvalues(a) and eigenfunctions(b) of $\tilde{\chi}$ and $\tilde{\chi}_{\text{RPA}}$ obtained from finite-field calculations. In (b), the first $32 \times 32$ elements of the $\braket{\xi^{\text{RPA}} | \xi^{\text{full}}}$ matrices are presented.}
  \label{chi_vs_chirpa}
\end{figure}

Figure \ref{chi_vs_chirpa} shows the eigendecomposition of $\tilde{\chi}$ compared to that of $\tilde{\chi}_{\text{RPA}}$.

As indicated by Figure \ref{chi_vs_chirpa}a, including $f_{\text{xc}}$ in the evaluation of $\chi$ results in a stronger screening. The eigenvalues of $\tilde{\chi}$ are systematically more negative than those of $\tilde{\chi}_{\text{RPA}}$, though they asymptotically converge to zero in the same manner. While the eigenvalues are different, the eigenvectors (eigenspaces in the case of degenerate eigenvalues) are almost identical, as indicated by the block-diagonal form of the eigenvector overlap matrices (see Figure \ref{chi_vs_chirpa}b). 

\begin{figure}[H]
  \includegraphics[width=3.33in]{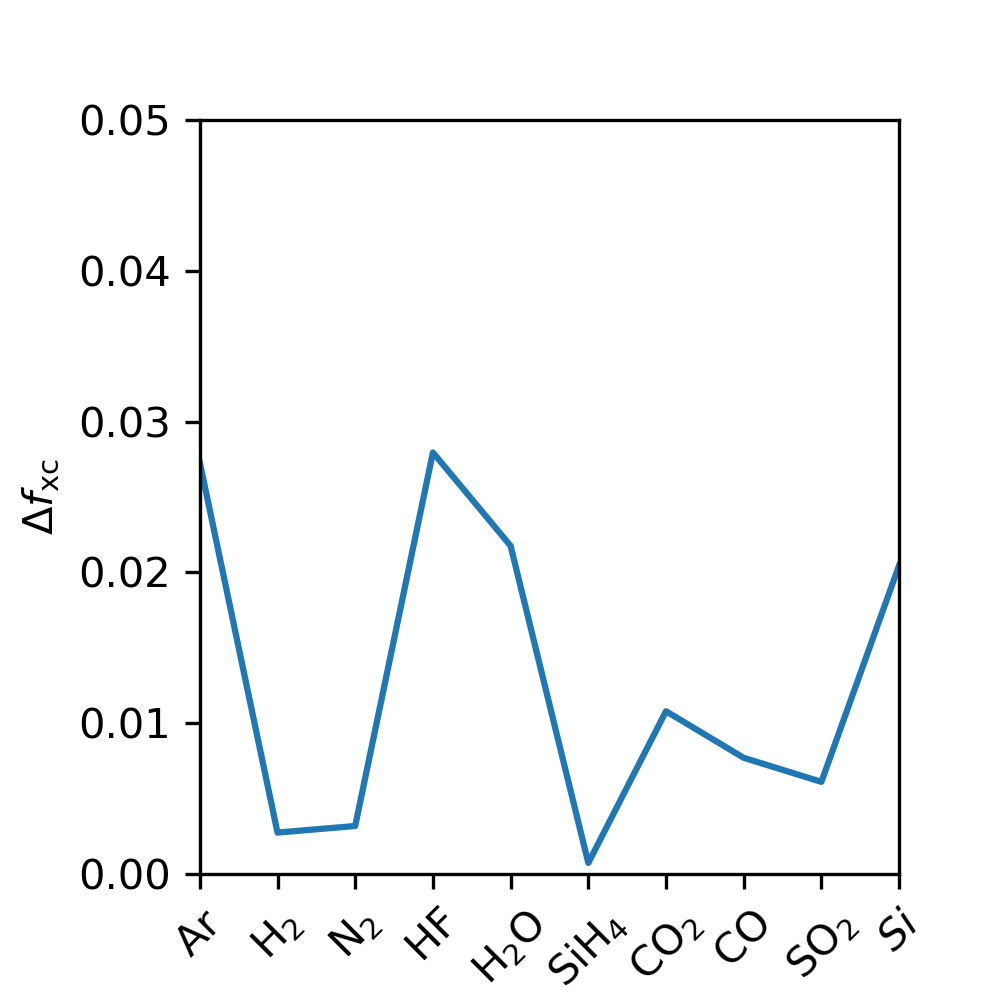}
  \caption{Average relative differences $\Delta f_{\text{xc}}$ (see eq \ref{deltafxc}) between diagonal elements of the $\tilde{f}_{\text{xc}}$ matrices computed analytically and numerically with the finite-field approach. Calculations were performed with the LDA functional.}
  \label{fxc_lda}
\end{figure}

Finally, $\tilde{f}_{\text{xc}}$ can be computed from $\tilde{\chi}$ and $\tilde{\chi}_0$ according to eq \ref{fxcmatrix}. Due to the similarity of the eigenvectors of $\tilde{\chi}$ and $\tilde{\chi}_{\text{RPA}}$ (identical to that of $\tilde{\chi}_0$), the $\tilde{f}_{\text{xc}}$ matrix is almost diagonal. In Section 3 of the SI we show the $\tilde{f}_{\text{xc}}$ matrix in the PDEP basis for a few systems. To verify the accuracy of $\tilde{f}_{\text{xc}}$ obtained by the finite-field approach, we performed calculations with the LDA functional, for which $f_{\text{xc}}$ can be computed analytically. In Figure \ref{fxc_lda} we present for a number of systems the average relative difference of the diagonal terms of the $\tilde{f}_{\text{xc}}$ matrices obtained analytically and through finite-field (FF) calculations. We define $\Delta f_{\text{xc}}$ as
\begin{equation} \label{deltafxc}
    \Delta f_{\text{xc}} = \frac{1}{N_{\text{PDEP}}} \sum_i^{N_{\text{PDEP}}} \frac{\left| \mel{\xi_i}{\tilde{f}_{\text{xc}}^{\text{FF}}}{\xi_i} - \mel{\xi_i}{\tilde{f}_{\text{xc}}^{\text{analytical}}}{\xi_i} \right|}{\left| \mel{\xi_i}{\tilde{f}_{\text{xc}}^{\text{analytical}}}{\xi_i} \right|} .
\end{equation}

As shown in Figure \ref{fxc_lda}, $\Delta f_{\text{xc}}$ is smaller than a few percent for all systems studied here. To further quantify the effect of the small difference found for the $\tilde{f}_{\text{xc}}$ matrices on $GW$ quasiparticle energies, we performed $G_0W_0^{f_\text{xc}}@\text{LDA}$ calculations for all the systems shown in Figure \ref{fxc_lda}, using the analytical $f_{\text{xc}}$ and $f_{\text{xc}}$ computed from finite-field calculations. The two approaches yielded almost identical quasiparticle energies, with mean absolute deviations of 0.04 and 0.004 eV for HOMO and LUMO levels, respectively.

\section{$GW$ calculations}

\subsection{Formalism}

In this section we discuss $GW$ calculations within and beyond the RPA, utilizing $f_{\text{xc}}$ computed with the finite-field approach. In the following equations we use 1, 2, ... as shorthand notations for $(\bm{r}_1, t_1)$, $(\bm{r}_2, t_2)$, ... Indices with bars are integrated over. When no indices are shown, the equation is a matrix equation in reciprocal space or in the PDEP basis. The following discussion focuses on finite systems; for periodic systems a special treatment of the long-range limit of $\chi$ is required and relevant formulae are presented in Section 4 of the SI.

Based on a KS reference system, the Hedin equations \cite{Hedin1965} relate the exchange-correlation self-energy $\Sigma_{\text{xc}}$ (abbreviated as $\Sigma$), Green's function $G$, the screened Coulomb interaction $W$, the vertex $\Gamma$ and the irreducible polarizability $P$:
\begin{equation} \label{Hedin1}
    \Sigma(1, 2) = i G(1, \bar4) W(1^+, \bar3) \Gamma(\bar4, 2; \bar3),
\end{equation}
\begin{equation} \label{Hedin2}
    W(1, 2) = v_{\text{c}}(1, 2) + v_{\text{c}}(1, \bar3) P(\bar3, \bar4) W(\bar4, 2),
\end{equation}
\begin{equation} \label{Hedin3}
    P(1, 2) = -i G(1, \bar3) G(\bar4, 1) \Gamma(\bar3, \bar4, 2),
\end{equation}
\begin{equation} \label{Hedin4}
    \Gamma(1, 2; 3) = \delta(1, 2) \delta(1, 3) + \frac{\delta \Sigma(1, 2)}{\delta G(\bar4, \bar5)} G(\bar4, \bar6) G(\bar7, \bar5) \Gamma(\bar6, \bar7, 3),
\end{equation}
\begin{equation} \label{Hedin5}
    G(1, 2) = G^0(1, 2) + G^0(1, \bar3) \Sigma(\bar3, \bar4) G(\bar4, 2).
\end{equation}

We consider three different $G_0W_0$ approximations: the first is the common $G_0W_0$ formulation within the RPA, here denoted as $G_0W_0^{\text{RPA}}$, where $\Gamma(1, 2; 3) = \delta(1, 2) \delta(1, 3)$ and $\Sigma$ is given by:
\begin{equation}
    \Sigma(1, 2) = i G(1, 2) W_{\text{RPA}}(1^+, 2),
\end{equation}
where
\begin{equation} \label{wrpa}
    W_{\text{RPA}}(1, 2) = v_{\text{c}}(1, 2) + v_{\text{c}}(1, \bar3) \chi_{\text{RPA}}(\bar3, \bar4) v_{\text{c}}(\bar4, 2),
\end{equation}
and
\begin{equation} \label{chirpa}
    \chi_{\text{RPA}} = (1 - \chi_0 v_{\text{c}})^{-1} \chi_0.
\end{equation}

The second approximation, denoted as $G_0W_0^{f_\text{xc}}$, includes $f_{\text{xc}}$ in the definition of $W$. Specifically, $\chi$ is computed from $\chi_0$ and $f_{\text{xc}}$ with eq \ref{Dyson}:
\begin{equation} \label{chifxc}
    \chi = (1 - \chi_0 (v_c + f_{\text{xc}}))^{-1} \chi_0,
\end{equation}
and is used to construct the screened Coulomb interaction beyond the RPA:
\begin{equation} \label{wfxc}
W_{f_\text{xc}} = v_{\text{c}}(1, 2) + v_{\text{c}}(1, \bar3) \chi(\bar3, \bar4) v_{\text{c}}(\bar4, 2).
\end{equation}

The third approximation, denoted as $G_0W_0\Gamma_0$, includes $f_{\text{xc}}$ in both $W$ and $\Sigma$. In particular, an initial guess for $\Sigma$ is constructed from $V_{\text{xc}}$:
\begin{equation} \label{sigmaxc}
    \Sigma_0(1, 2) = \delta(1, 2) V_{\text{xc}}(1)
\end{equation}
from which one can obtain a zeroth order vertex function by iterating Hedin's equations once \cite{DelSole1994}:
\begin{equation}
    \begin{split}
        \Gamma_0(1, 2; 3) = \delta(1, 2) (1 - f_{\text{xc}}\chi_0)^{-1}(1, 3).
    \end{split}
\end{equation}

Then the self-energy $\Sigma$ is constructed using $G$, $W_{f_\text{xc}}$ and $\Gamma_0$:
\begin{equation} \label{Hedin1vertex}
\begin{split}
  \Sigma(1, 2)
    & = i G(1, \bar4) W_{f_\text{xc}}(1^+, \bar3) \Gamma_0(\bar4, 2; \bar3) \\
    & = i G(1, 2) W_{\Gamma}(1^+, \bar3)
\end{split}
\end{equation}
where we defined an effective screened Coulomb interaction\bibnote{One may note that $\tilde{\chi}_{\Gamma}$ is not symmetric with respect to its two indices, and it can be symmetrized by using $\tilde{\chi}_0 \tilde{f}_{\text{xc}} \rightarrow \frac{1}{2}(\tilde{\chi}_0 \tilde{f}_{\text{xc}} + \tilde{f}_{\text{xc}} \tilde{\chi}_0 )$ in eq \ref{chitildegamma}. We found that the symmetrization has negligible effects on quasiparticle energies. We performed $G_0W_0^{f_\text{xc}}$ calculations for systems as shown in Figure \ref{fxc_lda} with either symmetrized or unsymmetrized $\tilde{\chi}_{\Gamma}$, the mean absolute deviations for HOMO and LUMO quasiparticle energies are 0.006 eV and 0.001 eV respectively.}
\begin{equation} \label{wgamma}
    W_{\Gamma} = v_{\text{c}}(1, 2) + v_{\text{c}}(1, \bar3) \chi_{\Gamma}(\bar3, \bar4) v_{\text{c}}(\bar4, 2),
\end{equation}
\begin{equation} \label{chigamma}
    \chi_{\Gamma} = [v_{\text{c}} - v_{\text{c}} \chi_0 (v_{\text{c}} + f_{\text{xc}})]^{-1} - v_{\text{c}}^{-1}.
\end{equation}

The symmetrized forms of the three different density response functions (reducible polarizabilities) defined in eq \ref{chirpa}, \ref{chifxc}, \ref{chigamma} are:
\begin{equation} \label{chitilderpa}
\tilde{\chi}_{\text{RPA}} = [1 - \tilde{\chi}_0]^{-1} \tilde{\chi}_0 
\end{equation}
\begin{equation} \label{chitildefxc}
\tilde{\chi} = [1 - \tilde{\chi}_0 (1 + \tilde{f}_{\text{xc}})]^{-1} \tilde{\chi}_0
\end{equation}
\begin{equation} \label{chitildegamma}
\tilde{\chi}_{\Gamma} = [1 - \tilde{\chi}_0 (1 + \tilde{f}_{\text{xc}})]^{-1} - 1
\end{equation}

Eqs. \ref{chitilderpa}-\ref{chitildegamma} have been implemented in the WEST code \cite{Govoni2015}. 

We note that finite-field calculations yield $\tilde{f}_{\text{xc}}$ matrices at zero frequency. Hence the results presented here correspond to calculations performed within the adiabatic approximation, as they neglect the frequency dependence of $\tilde{f}_{\text{xc}}$. An interesting future direction would be to compute frequency-dependent $\tilde{f}_{\text{xc}}$ by performing finite-field calculations using real-time time-dependent DFT (RT-TDDFT).

When using the $G_0W_0\Gamma_0$ formalism, the convergence of quasiparticle energies with respect to $N_\text{PDEP}$ turned out to be extremely challenging. As discussed in ref \citenum{Schmidt2017} the convergence problem originates from the incorrect short-range behavior of $\tilde{f}_{\text{xc}}$. In Section 3.2 below we describe a renormalization scheme of $\tilde{f}_{\text{xc}}$ that improves the convergence of $G_0W_0\Gamma_0$ results.

\subsection{Renormalization of $f_{\text{xc}}$}

Thygesen and co-workers \cite{Schmidt2017} showed that $G_0W_0\Gamma_0@\text{LDA}$ calculations with $f_{\text{xc}}$ computed at the LDA level exhibit poor convergence with respect to the number of unoccupied states and plane wave cutoff. We observed related convergence problems of $G_0W_0\Gamma_0$ quasiparticle energies as a function of $N_\text{PDEP}$, the size of the basis set used here to represent response functions (see Section 5 of the SI). In this section we describe a generalization of the $f_{\text{xc}}$ renormalization scheme proposed by Thygesen and co-workers \cite{Olsen2012, Olsen2013, Patrick2015} to overcome convergence issues.

The approach of ref \citenum{Schmidt2017} is based on the properties of the homogeneous electron gas (HEG). For an HEG with density $n$, $f^{\text{HEG}}_{\text{xc}}[n] (\bm{r}, \bm{r}')$ depends only on $(\bm{r} - \bm{r}')$ due to translational invariance, and therefore $f^{\text{HEG}}_{\text{xc}}[n]_{\bm{G}\bm{G}'}(\bm{q})$ is diagonal in reciprocal space. We denote the diagonal elements of $f^{\text{HEG}}_{\text{xc}}[n]_{\bm{G}\bm{G}'}(\bm{q})$ as $f^{\text{HEG}}_{\text{xc}}[n] (\bm{k})$ where $\bm{k} = \bm{q} + \bm{G}$. When using the LDA functional, the exchange kernel $f_x$ exactly cancels the Coulomb interaction $v_c$ at wavevector $k = 2k_F$ (the correlation kernel $f_c$ is small compared to $f_\text{x}$ for $k \geq 2k_F$), where $k_F$ is the Fermi wavevector. For $k \geq 2k_F$, $f^{\text{HEG-LDA}}_{\text{xc}}$ shows an incorrect asymptotic behavior, leading to an unphysical correlation hole \cite{Olsen2012, Olsen2013}. Hence Thygesen and co-workers introduced a renormalized LDA kernel $f^{\text{HEG-rLDA}}_{\text{xc}} (k)$ by setting $f^{\text{HEG-rLDA}}_{\text{xc}} (k) = f^{\text{HEG-LDA}}_{\text{xc}} (k)$ for $k \leq 2k_F$ and $f^{\text{HEG-rLDA}}_{\text{xc}} (k) = - v_c (k)$ for $k > 2k_F$. They demonstrated that the renormalized $f_\text{xc}$ improves the description of the short-range correlation hole as well as the correlation energy, and when applied to $GW$ calculations substantially accelerates the basis set convergence of $G_0W_0\Gamma_0$ quasiparticle energies.

While within LDA $f_\text{xc}$ can be computed analytically and $ v_c + f_\text{x} = 0$ at exactly $k = 2k_F$, for a general functional it is not known \textit{a priori} at which $k$ this condition is satisfied. In addition, for inhomogenous systems such as molecules and solids the $f_{\text{xc}}$ matrix is not diagonal in reciprocal space. The authors of Ref \citenum{Schmidt2017} used a wavevector symmetrization approach to evaluate $f^{\text{HEG-rLDA}}_{\text{xc}}$ for inhomogenous systems, which is not easily generalizable to the formalism adopted in this work, where $f_{\text{xc}}$ is represented in the PDEP basis.

To overcome these difficulties, here we first diagonalize the $\tilde{f}_{\text{xc}}$ matrix in the PDEP basis: 
\begin{equation} \label{fxcraw}
\tilde{f}_{\text{xc}} = \sum_i^{N_\text{PDEP}} f_i \ket{\zeta_i}\bra{\zeta_i},
\end{equation}
where $f$ and $\zeta$ are eigenvalues and eigenvectors of $\tilde{f}_{\text{xc}}$. Then we define a renormalized $\tilde{f}_{\text{xc}}$ as: 
\begin{equation} \label{fxcrenormalized}
\tilde{f}_{\text{xc}}^r = \sum_i^{N_\text{PDEP}} \max(f_i, -1) \ket{\zeta_i}\bra{\zeta_i}.
\end{equation}

Note that for $\tilde{f}_{\text{xc}} = -1$, $f_{\text{xc}} = - v_c$, therefore $f_{\text{xc}}^r$ is strictly greater or equal to $-v_c$. When applied to the HEG, the $f_{\text{xc}}^r@\text{LDA}$ is equivalent to $f^{\text{HEG-rLDA}}_{\text{xc}}$ in the limit $N_{\text{PDEP}} \rightarrow \infty$, where the PDEP and plane-wave basis are related by a unitary transformation. Thus, eq \ref{fxcrenormalized} represents a generalization of the scheme of Thygesen \textit{et al.} to any functional and to inhomogeneous electron gases. When using $f_{\text{xc}}^r$, we observed a faster basis set convergence of $G_0W_0\Gamma_0$ results than $G_0W_0^\text{RPA}$ results, consistent with ref \citenum{Schmidt2017}. In Section 5 of the SI we discuss in detail the effect of the $f_{\text{xc}}$ renormalization on the description of the density response functions $\chi$ and $\chi_{\Gamma}$, and we rationalize why the renormalization improves the convergence of $G_0W_0\Gamma_0$ results. Here we only mention that the response function $\tilde{\chi}_\Gamma$ may possess positive eigenvalues for large PDEP indices. When the renormalized $f_{\text{xc}}$ is used, the eigenvalues of $\tilde{\chi}_\Gamma$ are guaranteed to be nonpositive and they decay rapidly toward zero as the PDEP index increase, which explains the improved convergence of $G_0W_0\Gamma_0$ quasiparticle energies.

All $G_0W_0\Gamma_0$ results shown in Section 3.3 were obtained with renormalized $f_\text{xc}$ matrices, while $G_0W_0^{f_\text{xc}}$ calculations were performed without renormalizing $f_\text{xc}$, since we found that the renormalization had a negligible effect on $G_0W_0^{f_\text{xc}}$ quasiparticle energies (see SI Section 5).

\subsection{Results}

In this section we report $GW$ quasiparticle energies for molecules in the GW100 set \cite{vanSetten2015} and for several solids. Calculations are performed at $G_0W_0^{\text{RPA}}$, $G_0W_0^{f_\text{xc}}$ and $G_0W_0\Gamma_0$ levels of theory and with semilocal and hybrid functionals. Computational parameters including $E_\text{cut}$ and $N_\text{PDEP}$ for all calculations are summarized in Section 1 of the SI. A discussion of the convergence of $G_0W_0^{\text{RPA}}$ quasiparticle energies with respect to these parameters can be found in ref \citenum{Govoni2018}.

\begin{figure}[H]
  \centering
  \includegraphics[width=5in]{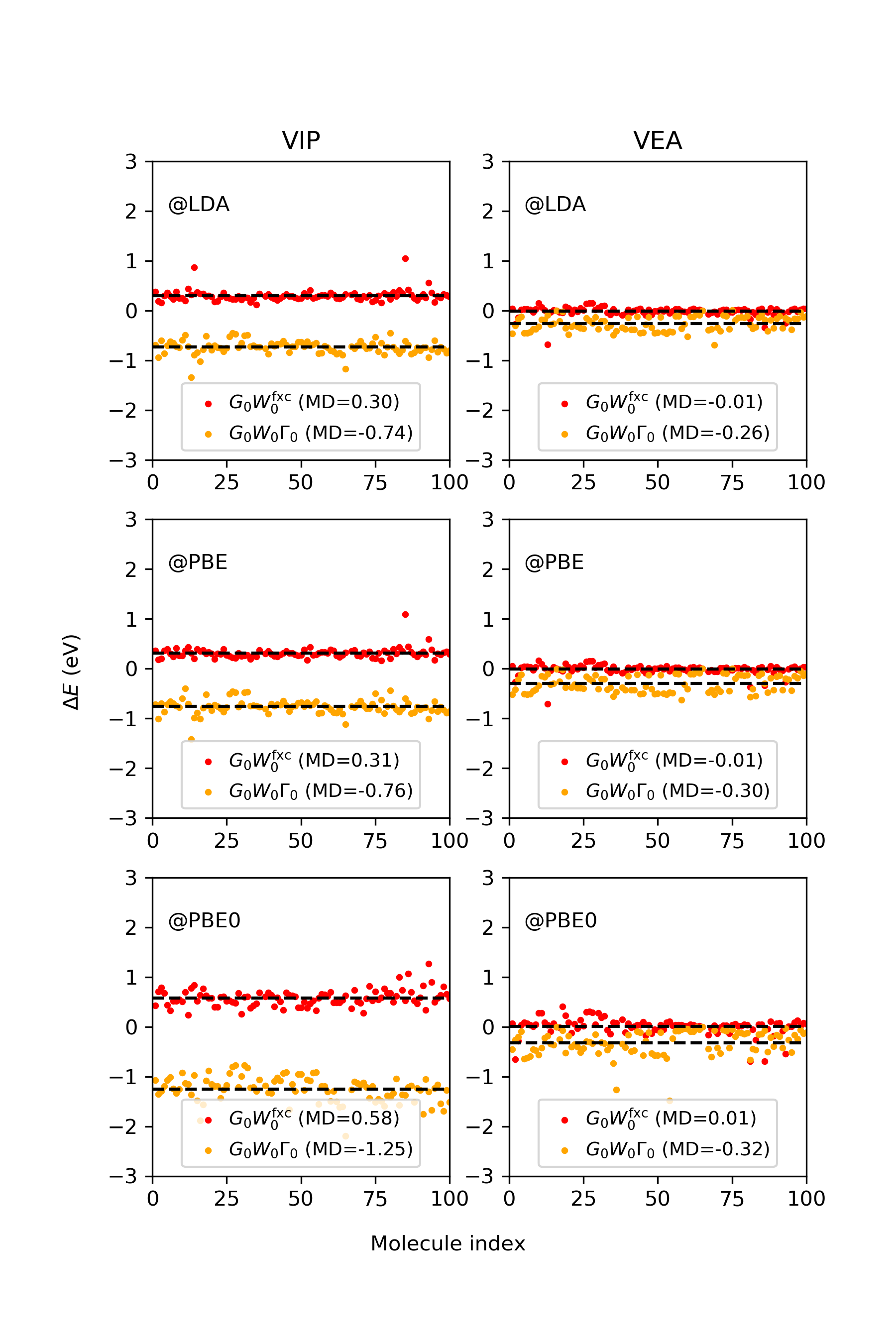}
  \caption{Difference ($\Delta E$) between vertical ionization potential (VIP) and vertical electron affinity (VEA) of molecules in the GW100 set computed at the $G_0W_0^{f_\text{xc}}$/$G_0W_0\Gamma_0$ level and corresponding $G_0W_0^{\text{RPA}}$ results. Mean deviations (MD) in eV are shown in brackets and represented with black dashed lines. Results are presented for three different functionals (LDA, PBE and PBE0) in the top, middle and bottom panel, respectively. }
  \label{gw100_gw_compare}
\end{figure}

We computed the vertical ionization potential (VIP), vertical electron affinity (VEA) and fundamental gaps for molecules with LDA, PBE and PBE0 functionals. VIP and VEA are defined as $\text{VIP} = \varepsilon^{\text{vac}} - \varepsilon^{\text{HOMO}}$ and $\text{VEA} = \varepsilon^{\text{vac}} - \varepsilon^{\text{LUMO}}$ respectively, where $\varepsilon^{\text{vac}}$ is the vacuum level estimated with the Makov-Payne method \cite{Makov1995}; $\varepsilon^{\text{HOMO}}$ and $\varepsilon^{\text{LUMO}}$ are HOMO and LUMO $GW$ quasiparticle energies, respectively. The results are summarized in Figure \ref{gw100_gw_compare}, where VIP and VEA computed at $G_0W_0^{f_\text{xc}}$ and $G_0W_0\Gamma_0$ levels are compared to results obtained at the $G_0W_0^{\text{RPA}}$ level \bibnote{For \ce{KH} molecule, $G_0W_0^{f_\text{xc}}@\text{PBE}$ calculation for the HOMO converged to a satellite instead of the quasiparticle peak. The spectral function of \ce{KH} is plotted and discussed in SI Section 6 and the correct quasiparticle energy is used here.}.

Compared to $G_0W_0^{\text{RPA}}$ results, the VIP computed at the $G_0W_0^{f_\text{xc}}$/$G_0W_0\Gamma_0$ level are systematically higher/lower, and the deviation of $G_0W_0\Gamma_0$ from $G_0W_0^{\text{RPA}}$ results is more than twice as large as that of $G_0W_0^{f_\text{xc}}$ results. The differences reported in Figure \ref{gw100_gw_compare} are more significant with hybrid functional starting point, as indicated by the large mean deviations (MD) for $G_0W_0\Gamma_0$/$G_0W_0\Gamma_0$ results obtained with the PBE0 functional (0.58/-1.25 eV) compared to the MD of semilocal functionals (0.30/-0.74 eV for LDA and 0.31/-0.76 eV for PBE). In contrast to VIP, VEA appear to be less affected by vertex corrections. $G_0W_0^{f_\text{xc}}$ does not systematically shift the VEA from $G_0W_0^{\text{RPA}}$ results. $G_0W_0\Gamma_0$ calculations result in systematically lower VEA than those obtained at the $G_0W_0^{\text{RPA}}$ level by about 0.3 eV with all DFT starting points, but overall the deviations are much smaller than for the VIP.

\begin{figure}[H]
  \centering
  \includegraphics[width=7in]{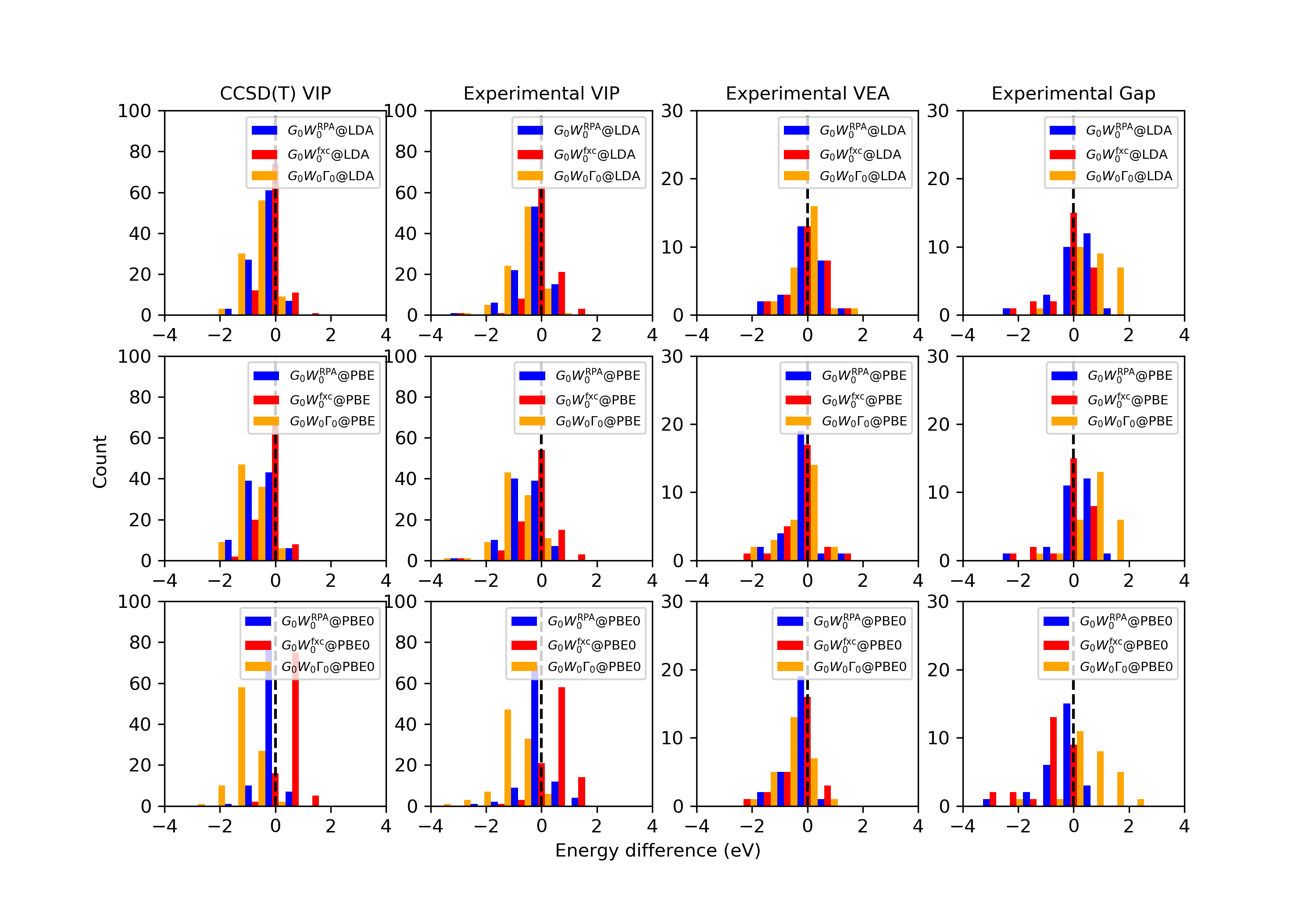}
  \caption{Vertical ionization potential (VIP), vertical electron affinity (VEA) and electronic gap of molecules in the GW100 set computed at $G_0W_0^{\text{RPA}}$, $G_0W_0^{f_\text{xc}}$ and $G_0W_0\Gamma_0$ levels of theory, compared to experimental and CCSD(T) results (black dashed lines). }
  \label{gw100_ref_compare}
\end{figure}

In Figure \ref{gw100_ref_compare} we compare $GW$ results with experiments \bibnote{WEST GW100 data collection. http://www.west-code.org/database (accessed Aug. 1, 2018).} and quantum chemistry CCSD(T) results \cite{Krause2015}. The corresponding MD and mean absolute deviations (MAD) are summarized in Table \ref{gw100_ref_deviation}. At the $G_0W_0^{\text{RPA}}@\text{PBE}$ level, the MAD for the computed VIP values compared to CCSD(T) and experimental results are 0.50 and 0.55 eV respectively, and the MAD for the computed VEA compared to experiments is 0.46 eV. These MAD values (0.50/0.55/0.46 eV) are comparable to previous benchmark studies on the GW100 set using the FHI-aims (0.41/0.46/0.45 eV) \cite{vanSetten2015}, VASP (0.44/0.49/0.42 eV) \cite{Maggio2017} and WEST codes (0.42/0.46/0.42 eV) \cite{Govoni2018}, although in this work we did not extrapolate our results with respect to the basis set due to the high computational cost. 

Compared to experiments and CCSD(T) results, $G_0W_0^{f_\text{xc}}$ improves over $G_0W_0^{\text{RPA}}$ for the calculation of VIP when semilocal functional starting points (LDA, PBE) are used, as indicated by the values of MD and MAD of $G_0W_0^{f_\text{xc}}@\text{LDA}/\text{PBE}$ results compared to that of $G_0W_0^{\text{RPA}}@\text{LDA}/\text{PBE}$. When using the PBE0 functional as starting point, $G_0W_0^{f_\text{xc}}$ leads to an overestimation of VIP by 0.53 eV on average. $G_0W_0\Gamma_0$ calculations underestimate VIP by about 1 eV with all functionals tested here. For the calculation of VEA, $G_0W_0^{f_\text{xc}}$ performs similarly to $G_0W_0^{\text{RPA}}$ as discussed above, and $G_0W_0\Gamma_0$ yields an underestimation of 0.25/0.43/0.64 eV on average with LDA/PBE/PBE0 starting points compared to experiments.

\begin{table}[H]
\caption{Mean deviation and mean absolute deviation (in brackets) for $GW$ results compared to experimental results and CCSD(T) calculations. We report vertical ionization potentials (VIP), vertical electron affinities (VEA) and the fundamental electronic gaps. All values are given in eV.}
\begin{tabular}{llrrrr}
\hline
{} &   CCSD(T) VIP & Exp. VIP & Exp. VEA & Exp. Gap \\
\hline
$G_0W_0^{\mathrm{RPA}}@\mathrm{LDA}$  &  -0.23 (0.34) &     -0.19 (0.43) &      0.04 (0.45) &      0.21 (0.56) \\
$G_0W_0^{f_\mathrm{xc}}@\mathrm{LDA}$  &   0.06 (0.29) &      0.11 (0.37) &      0.03 (0.48) &     -0.10 (0.49) \\
$G_0W_0\Gamma_0@\mathrm{LDA}$         &  -0.97 (0.98) &     -0.93 (0.95) &     -0.25 (0.41) &      0.59 (0.75) \\
\hline
$G_0W_0^{\mathrm{RPA}}@\mathrm{PBE}$  &  -0.43 (0.50) &     -0.39 (0.55) &     -0.09 (0.46) &      0.28 (0.57) \\
$G_0W_0^{f_\mathrm{xc}}@\mathrm{PBE}$  &  -0.12 (0.32) &     -0.07 (0.43) &     -0.10 (0.49) &     -0.05 (0.46) \\
$G_0W_0\Gamma_0@\mathrm{PBE}$         &  -1.19 (1.20) &     -1.15 (1.16) &     -0.43 (0.53) &      0.64 (0.79) \\
\hline
$G_0W_0^{\mathrm{RPA}}@\mathrm{PBE0}$ &  -0.05 (0.20) &     -0.01 (0.34) &     -0.26 (0.41) &     -0.26 (0.47) \\
$G_0W_0^{f_\mathrm{xc}}@\mathrm{PBE0}$ &   0.53 (0.57) &      0.57 (0.65) &     -0.27 (0.50) &     -0.83 (0.83) \\
$G_0W_0\Gamma_0@\mathrm{PBE0}$        &  -1.30 (1.30) &     -1.26 (1.26) &     -0.64 (0.68) &      0.50 (0.72) \\
\hline
\end{tabular}
\label{gw100_ref_deviation}
\end{table}

Finally we report $G_0W_0^{\text{RPA}}$, $G_0W_0^{f_\text{xc}}$ and $G_0W_0\Gamma_0$ results for several solids: \ce{Si}, \ce{SiC} (4H), \ce{C} (diamond), \ce{AlN}, \ce{WO3} (monoclinic), \ce{Si3N4} (amorphous). We performed calculations starting with LDA and PBE functionals for all solids, and for Si we also performed calculations with a dielectric-dependent hybrid (DDH) functional \cite{Skone2014}. All solids are represented by supercells with 64-96 atoms (see Section 1 of the SI) and only the $\Gamma$-point is used to sample the Brillioun zone. In Table \ref{bandgaps} we present the band gaps computed with different $GW$ approximations and functionals. Note that the supercells used here do not yield fully converged results as a function of supercell size (or k-point sampling); however the comparisons between different $GW$ calculations are sound and represent the main result of this section.

\begin{table}[H]
\caption{Band gaps (eV) for solids computed by different $GW$ approximations and exchange-correlation (XC) functionals (see text). All calculations are performed at the $\Gamma$-point of supercells with 64-96 atoms (see Section 1 of the SI for details). }
\begin{tabular}{llrrrr}
\hline
    &     &  DFT & $G_0W_0^{\mathrm{RPA}}$ & $G_0W_0^{f_\mathrm{xc}}$ & $G_0W_0\Gamma_0$ \\
System & XC &      &                         &                         &                           \\
\hline
\ce{Si} & LDA & 0.55 &                    1.35 &                     1.33 &             1.24 \\
      & PBE & 0.73 &                    1.39 &                     1.37 &             1.28 \\
      & DDH & 1.19 &                    1.57 &                     1.50 &             1.48 \\
\hline
\ce{C} (diamond) & LDA & 4.28 &                    5.99 &                     6.00 &             5.89 \\
      & PBE & 4.46 &                    6.05 &                     6.06 &             5.95 \\
\hline
\ce{SiC} (4H) & LDA & 2.03 &                    3.27 &                     3.23 &             3.26 \\
      & PBE & 2.21 &                    3.28 &                     3.23 &             3.28 \\
\hline
\ce{AlN} & LDA & 3.85 &                    5.67 &                     5.72 &             5.66 \\
      & PBE & 4.04 &                    5.67 &                     5.74 &             5.68 \\
\hline
\ce{WO3} (monoclinic) & LDA & 1.68 &                    3.10 &                     3.07 &             3.15 \\
      & PBE & 1.78 &                    2.97 &                     2.87 &             3.03 \\
\hline
\ce{Si3N4} (amorphous) & LDA & 3.04 &                    4.84 &                     4.92 &             4.81 \\
      & PBE & 3.19 &                    4.86 &                     4.96 &             4.83 \\
\hline
\end{tabular}
\label{bandgaps}
\end{table}

\begin{figure}[H]
  \centering
  \includegraphics[width=3.33in]{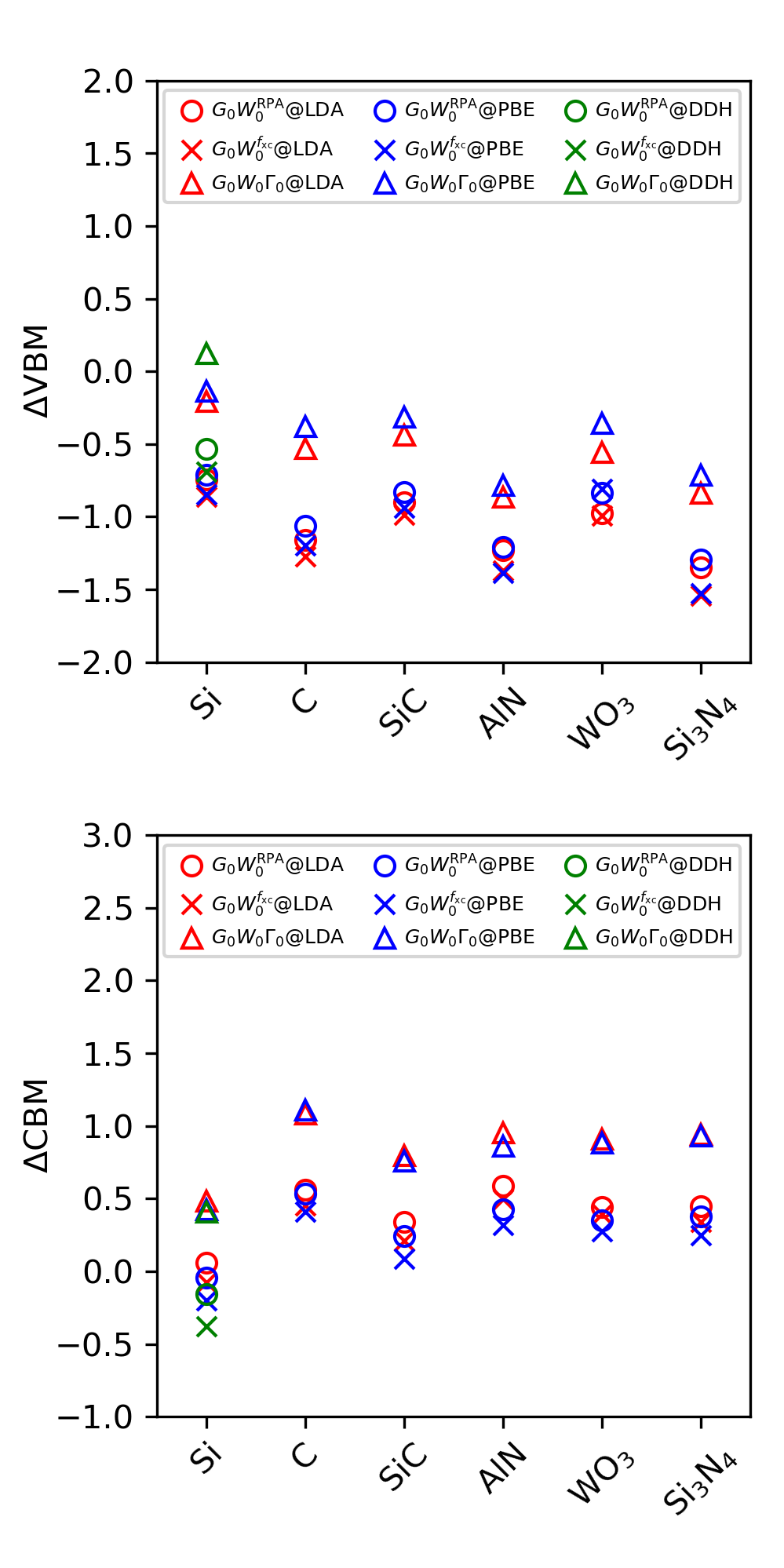}
  \caption{$GW$ quasiparticle corrections to the valance band maximum (VBM) and the conduction band minimum (CBM). Circles, squares and triangles are $G_0W_0^{\text{RPA}}$, $G_0W_0^{f_\text{xc}}$ and $G_0W_0\Gamma_0$ results respectively; red, blue, green markers correspond to calculations with LDA, PBE and DDH functionals.}
  \label{bandedges_solids}
\end{figure}

Overall, band gaps obtained with different $GW$ approximations are rather similar, with differences much smaller than those observed for molecules. To further investigate the positions of the band edges obtained from different $GW$ approximations, we plotted in Figure \ref{bandedges_solids} the $GW$ quasiparticle corrections to VBM and CBM, defined as $\Delta_{\mathrm{VBM/CBM}} = \varepsilon^{\text{GW}}_{\text{VBM/CBM}} - \varepsilon^{\text{DFT}}_{\text{VBM/CBM}}$ where $\varepsilon^{\text{GW}}_{\text{VBM/CBM}}$ and $\varepsilon^{\text{DFT}}_{\text{VBM/CBM}}$ are the $GW$ quasiparticle energy and the Kohn-Sham eigenvalue corresponding to the VBM/CBM, respectively.

Compared to $G_0W_0^{\text{RPA}}$, VBM and CBM computed at the $G_0W_0^{f_\text{xc}}$ level are slightly lower, while VBM and CBM computed at the $G_0W_0\Gamma_0$ level are significantly higher. For Si, $\Delta_{\mathrm{VBM/CBM}}$ obtained with LDA starting points are -0.75/0.06 ($G_0W_0^{\text{RPA}}$), -0.86/-0.08 ($G_0W_0^{f_\text{xc}}$), -0.21/0.49 ($G_0W_0\Gamma_0$) eV respectively, showing a trend in agreement with the results reported by Del Sole \textit{et al} (-0.36/0.27, -0.44/0.14, 0.01/0.67 eV) \cite{DelSole1994}, but with an overall overestimate of the band gap due a lack of convergence in our Brillouin zone sampling. The difference between band edge energies computed by different $GW$ approximations is larger with the DDH functional, compared to that of semilocal functionals. Overall the trends observed for solids are consistent with those found for molecules, except that for solids the shift of the CBM resembles those of the VBM when vertex corrections are included, while for molecules VEA is less sensitive to vertex corrections.

\section{Conclusions}
In summary, we developed a finite-field approach to compute density response functions ($\chi_0$, $\chi_{\text{RPA}}$ and $\chi$) for molecules and materials. The approach is non-perturbative and can be used in a straightforward manner with both semilocal and orbital-dependent functionals. Using this approach, we computed the exchange-correlation kernel $f_{\text{xc}}$ and performed $GW$ calculations using dielectric responses evaluated beyond the RPA.

We evaluated quasiparticle energies for molecules and solids and compared results obtained within and beyond the RPA, and using DFT calculations with semilocal and hybrid functionals as input. We found that the effect of vertex corrections on quasiparticle energies is more notable when using input wavefunctions and single-particle energies from hybrid functionals calculations. For the small molecules in the GW100 set, $G_0W_0^{f_\text{xc}}$ calculations yielded higher VIP compared to $G_0W_0^{\text{RPA}}$ results, leading to a better agreement with experimental and high-level quantum chemistry results when using LDA and PBE starting points, and to a slight overestimate of VIP when using PBE0 as the starting point. $G_0W_0\Gamma_0$ calculations instead yielded a systematic underestimate of VIP of molecules. VEA of molecules were found to be less sensitive to vertex corrections compared to VIP. In the case of solids, the energy of the VBM and CBM shifts in the same direction, relative to RPA results, when vertex corrections are included, and overall the band gaps were found to be rather insensitive to the choice of the $GW$ approximation.

In addition, we reported a scheme to renormalize $f_{\text{xc}}$, which is built on previous work \cite{Schmidt2017} using the LDA functional. The scheme is general and applicable to any exchange-correlation functional and to inhomogeneous systems including molecules and solids. Using the renormalized $\tilde{f}_{\text{xc}}$, the basis set convergence of $G_0W_0\Gamma_0$ results was significantly improved.

Overall, the method introduced in our work represents a substantial progress towards efficient computations of dielectric screening and large-scale $G_0W_0$ calculations for molecules and materials beyond the random phase approximation.

\acknowledgement
We thank Timothy Berkelbach, Alan Lewis and Ngoc Linh Nguyen for helpful discussions. This work was supported by MICCoM, as part of the Computational Materials Sciences Program funded by the U.S. Department of Energy, Office of Science, Basic Energy Sciences, Materials Sciences and Engineering Division through Argonne National Laboratory, under contract number DE-AC02-06CH11357. This research used resources of the National Energy Research Scientific Computing Center (NERSC), a DOE Office of Science User Facility supported by the Office of Science of the US Department of Energy under Contract No. DE-AC02-05CH11231, resources of the Argonne Leadership Computing Facility, which is a DOE Office of Science User Facility supported under Contract No. DE-AC02-06CH11357, and resources of the University of Chicago Research Computing Center.

\suppinfo
The Supporting Information contains parameters used for calculations, convergence tests, detailed discussion of $\tilde{f}_\text{xc}$ matrix and its renormalization, extension of beyond-RPA $GW$ formalism to solids, and an analysis of the spectral function of \ce{KH} molecule.

\pagebreak
Table of Contents:
\begin{figure}[H]
  \includegraphics[width=7in]{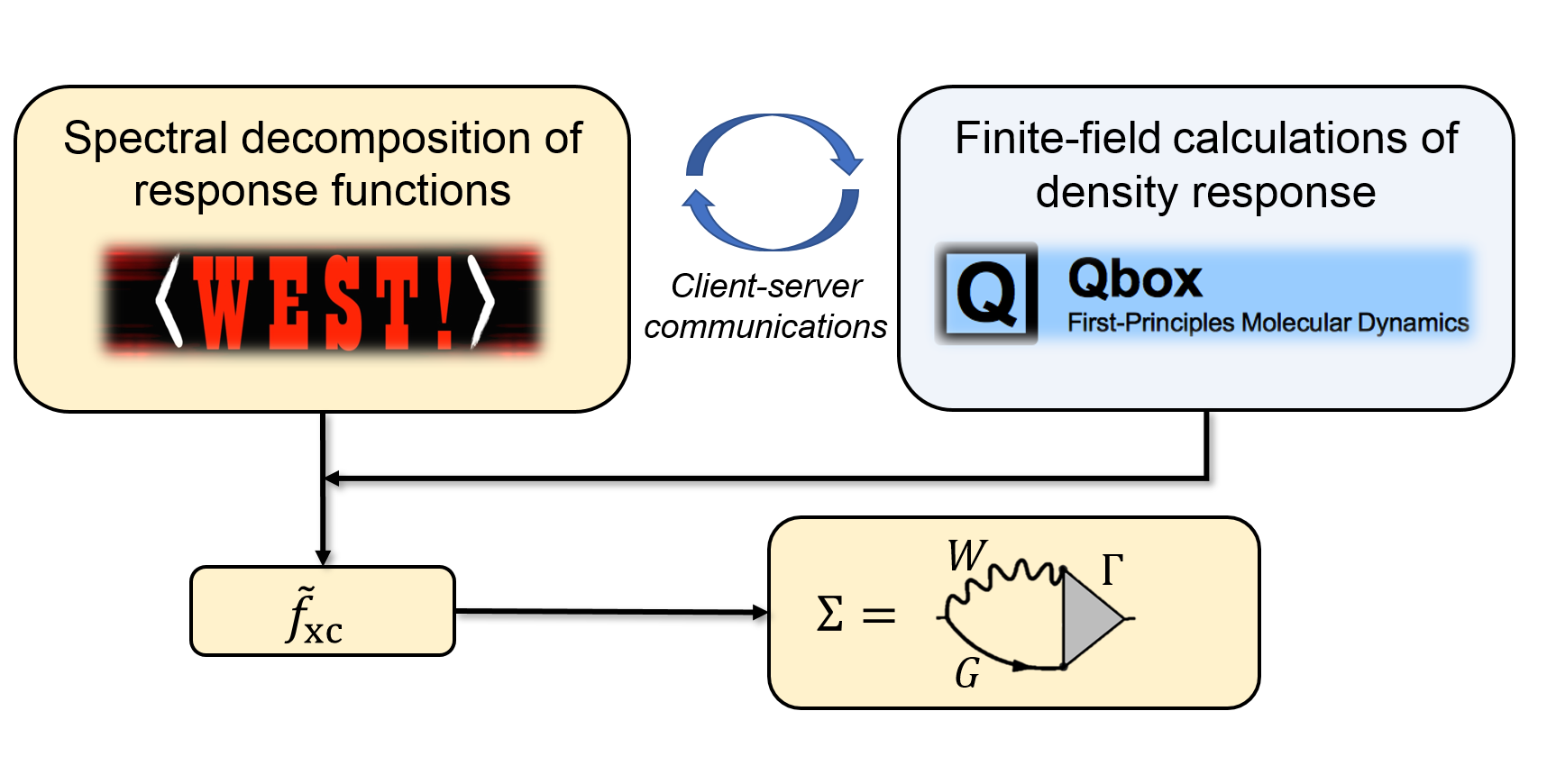}
\end{figure}

\pagebreak
\bibliography{references}

\begin{mcitethebibliography}{74}
\bibitem[Not()]{Note-3}
Here, we defined the PDEP basis to be the eigenvectors of $\tilde{\chi}_{0}$.
  Alternatively, one may first iteratively diagonalize $\tilde{\chi}$ and
  define its eigenvectors as the PDEP basis. Then $\tilde{\chi}_0$ and
  $\tilde{f}_{\text{xc}}$ can be evaluated in the space of the eigenvectors of
  $\tilde{\chi}$. This choice is not further discussed in the paper; we only
  mention that some comparisons for the quasiparticle energies (at the
  $G_0W_0^{f_\text{xc}}$ level, see Section 3) of selected molecules obtained
  using either $\tilde{\chi}_0$ or $\tilde{\chi}$ eigenvectors as the PDEP
  basis are identical within 0.01 (0.005) eV for the HOMO (LUMO) state.\relax
\mciteBstWouldAddEndPunctfalse
\mciteSetBstMidEndSepPunct{\mcitedefaultmidpunct}
{}{\mcitedefaultseppunct}\relax
\EndOfBibitem
\bibitem[Not()]{Note-5}
One may note that $\tilde{\chi}_{\Gamma}$ is not symmetric with respect to its
  two indices, and it can be symmetrized by using $\tilde{\chi}_0
  \tilde{f}_{\text{xc}} \rightarrow \frac{1}{2}(\tilde{\chi}_0
  \tilde{f}_{\text{xc}} + \tilde{f}_{\text{xc}} \tilde{\chi}_0 )$ in eq
  \ref{chitildegamma}. We found that the symmetrization has negligible effects
  on quasiparticle energies. We performed $G_0W_0^{f_\text{xc}}$ calculations
  for systems as shown in Figure \ref{fxc_lda} with either symmetrized or
  unsymmetrized $\tilde{\chi}_{\Gamma}$, the mean absolute deviations for HOMO
  and LUMO quasiparticle energies are 0.006 eV and 0.001 eV respectively.\relax
\mciteBstWouldAddEndPunctfalse
\mciteSetBstMidEndSepPunct{\mcitedefaultmidpunct}
{}{\mcitedefaultseppunct}\relax
\EndOfBibitem
\bibitem[Not()]{Note-6}
For \ce{KH} molecule, $G_0W_0^{f_\text{xc}}@\text{PBE}$ calculation for the
  HOMO converged to a satellite instead of the quasiparticle peak. The spectral
  function of \ce{KH} is plotted and discussed in SI Section 6 and the correct
  quasiparticle energy is used here.\relax
\mciteBstWouldAddEndPunctfalse
\mciteSetBstMidEndSepPunct{\mcitedefaultmidpunct}
{}{\mcitedefaultseppunct}\relax
\EndOfBibitem
\end{mcitethebibliography}

\end{document}